# System-Level Genetic Codes Using a Transposable Element-Like Mechanism with Applications to Cancer

By John F. McGowan, Ph.D.
Desktop Video Expert Center
NASA Ames Research Center
Mail Stop 233-18
Moffett Field, CA 94035-1000
E-Mail: jmcgowan@mail.arc.nasa.gov
Telephone: (650) 604-0143

(3/28/00 5:07 PM)

## ABSTRACT

A system-level genetic code is a hypothetical genetic code that exclusively or preferentially codes systems of interacting coadapted parts. System-level genetic codes differ from part-level genetic codes in which each discrete part is coded independently. In general, a system-level genetic code requires coding discrete interacting parts such as organs or proteins in an interdependent way. Changing a single symbol or "gene" in a system-level genetic code affects two or more parts in a coordinated way. System-level genetic codes provide a plausible mechanism for evolution by leaps also known as systemic macromutations or hopeful monsters without invoking an intelligent agent, vitalistic forces, or new laws of physics. A system-level genetic code may explain the rapid appearance of new forms of life in the fossil record such as the Cambrian Explosion in which most invertebrate phyla appeared. A system-level genetic code may explain the origin of complex integrative systems of tightly coadapted proteins such as the blood coagulation cascade in which the removal or structural modification of any single part breaks the system. It is demonstrated that a simple system-level genetic code can, in principle, be implemented using biochemical mechanisms similar or identical to the mechanisms used by transposable elements, regulation of gene expression, and the adaptive vertebrate immune system. It is further suggested that cancer in the body and both normal and tumor cell lines cultured in vitro are caused by mutations of the system-level genetic code. An explanation why multi-cellular organisms retain the ability to revert to uni-cellularity and thus a tendency to develop cancer is presented.

1. INTRODUCTION

A system-level genetic code is a hypothetical genetic code that preferentially or exclusively represents systems of interacting coadapted parts. It differs from a part-level genetic code where individual parts such as structural proteins are coded independently. A mutation of a single symbol in a system-level genetic code can change two or more parts simultaneously in a coordinated way. This paper reviews various problems with standard evolutionary theory that suggest a system-level genetic code may exist. It is demonstrated that simple system-level genetic codes based on context-free grammars can be implemented using biochemical mechanisms similar or identical to those observed in transposable elements, regulation of gene expression, and the adaptive immune system of jawed vertebrates. Finally, it is argued that a system-level genetic code using a transposable element-like mechanism may provide a unified theory of cancer in the body and the development of immortal normal and tumor cell lines in vitro.

A simple toy, the jigsaw puzzle illustrates a system-level genetic code. A jigsaw puzzle is a complex integrative system in which the removal of any single piece or structure modification of any single piece breaks the puzzle. If the pieces of the puzzle are independently coded – for example as an ordered list of the vertices of the piece – then any single mutation and the vast majority of simultaneous multiple mutations breaks the puzzle. This independent coding of pieces is not how jigsaw puzzles are designed or created. Jigsaw puzzles are formed by cutting a parent piece into a set of matching pieces with a literal or figurative jigsaw. If the jigsaw puzzle is coded as the shape of the parent piece, for example an ordered list of the vertices of the parent piece, and the rules for cutting the parent piece, any single mutation will produce a valid jigsaw puzzle. A single mutation will change several pieces simultaneously in a coordinated fashion.

The two ways of coding a jigsaw puzzle are examples of representations. A representation is a way of describing a system, usually exactly. In general, there are many representations for a system. Although different representations describe the same system, sometimes a problem will be intractable in one representation and easy to solve in another representation. For complex systems of parts, there is a family of part-level representations in which each part is described independently. Part-level description or part-level code may be substituted for part-level representation. There is also a family of system-level representations in which parts are described in an interdependent fashion. A system-level representation will preferentially or exclusively represent parts that are correlated in some essential way necessary for the function of the system. System-level description or system-level code may be substituted for system-level representation. In the following discussion, it is important to realize that the system-level representation and the part-level representations are equivalent. The system-level representation can be transformed into the part-level representation and vice-versa. The equivalence of two representations is sometimes called duality by mathematicians and physicists.

A system-level genetic code is a hypothetical genetic code that preferentially or exclusively represents systems of tightly coadapted parts. In systems designed by humans, this typically involves coding the interface shared by two or more parts as a single symbol of the code and coding the properties of the parts that are independent of the interfaces as independent symbols of the code[1]. Some properties are shared by all parts. Some properties are shared by groups of parts. Some properties are unique to a single part. For example, a jigsaw puzzle can be made of wood or cardboard. This is usually a shared property of all pieces in the puzzle. Ideally, the code for the interface elements is defined so that any string of symbols codes a valid interface. Any single-step mutation of the interface element generates another valid interface element. A mutation of the interface element can simultaneously change two or more parts in the system. Similarly, a single mutation of a property shared by all parts will change all parts in the system simultaneously. A system-level genetic code combined with random variation and natural selection could evolve complex integrated systems without resort to scaffoldings or indirect routes in which the system purportedly evolves from other systems that perform other functions.

Several reasons for suspecting a system-level genetic code exist[2]. The fossil record contains numerous cases of the sudden appearance of new forms. Intermediate forms are frequently absent or unrecognized. There is a troubling lack of convincing intermediate forms between seemingly widely separated forms where many, many intermediates would be expected. The fossil record looks suspiciously like large-scale changes, saltations or systemic macromutations that would require several parts to change at once, have occurred on several occasions. For example, in the Cambrian explosion of about 600 million years ago, most of the invertebrate phyla appeared in the fossil record during a period of only 50 million years, possibly 10 million years by some estimates. Intermediates and plausible precursors to the different invertebrate phyla are absent. The Cambrian explosion suggests a discontinuous jump from primitive multi-cellular organisms to fully functional animals.

Systemic macromutations or saltations, the "hopeful monsters" of Richard Goldschmidt, are implausible without new physical phenomena or an intelligent agent if the parts of the living systems are independently coded[3]. The jump from primitive multi-cellular organisms to multi-cellular invertebrates that the Cambrian explosion suggests would require many simultaneous harmonious changes. A jump from a terrestrial rodent to a functional bat would require dozens of bones and muscles to change simultaneously. If the parts are independently coded then the probability of random variation producing these jumps is essentially zero.

A system-level genetic code permits a single-step mutation to change two or more parts harmoniously and can, in principle, leap major functional gaps, such as the gaps between phyla, in a single step. In the toy example of the jigsaw puzzle the addition of a new cutting rule might correspond to a major jump between widely separated species. A change in a pre-existing cutting rule might correspond to a small jump between closely related species. Although a mutation in a system-level genetic code always generates a system of coadapted parts, most of these systems will be negative or neutral mutations. A species may remain stable for a long time.

Major gaps such as those between phyla will rarely be spanned. If a major jump occurs, smaller more probable jumps will then populate the branch of the tree of life corresponding to the new phylum. Once a single species of a new phylum is created, there will be a rapid, in geological time, proliferation of species belonging to the new phylum. Thus periods of mass proliferation of new species such as the Cambrian explosion are expected immediately following a major leap across a functional gap – for example, the leap from primitive multi-cellular organisms to animals. Once a working example of a basic type appears, this example will diversify into many new species within that type much faster because the gaps are smaller and the probability of spanning a small gap is greater than spanning a large gap. Thus the major features of the fossil record, both stasis and the sudden appearance of new species and entire types, can be explained with a system-level genetic code.

Many examples of complex integrative biochemical systems exist[4]. These include the blood clotting cascade, the cilium, the bacterial flagella, and the immune system. The genetic code itself, the DNA molecules, the histone proteins that bind chromosomes together, and the molecular machinery that decodes the genetic code, such as the ribosomes, seems to be an integrated system in which all parts are needed for function and all parts are tightly coupled. These biochemical systems suggest the simultaneous or nearly simultaneous appearance of several coadapted proteins and, indeed, the nucleic acids.

Gene knockout studies in which a single coding gene is removed usually show that a given gene affects several different systems in apparently disjoint ways. The gene does not change all of the systems in a harmonious manner as a system-level genetic code mutation would - at least sometimes. Rather all of the systems that appear to use the protein coded by the gene break. This substantially reduces the likelihood of the largely hypothetical positive mutations presumed to drive evolution. Since a protein is frequently reused in multiple different systems or used to form several longer proteins used in different systems, in general a mutation must not only improve the protein for one system but preserve or improve its function in all systems that utilize the protein. *Or* all of the different systems that use the protein must change together when the constituent protein changes. These results also suggest a system-level genetic code.

The pattern of differences in the amino acid sequences in proteins and nucleotide base pair sequences in the DNA between different species, usually explained through the molecular clock hypothesis, is difficult to account for with the traditional part-level genetic code. Specifically, different species appear to have widely differing generation times and annual rates of mutation. The molecular clock hypothesis appears to require a constant rate of mutation per unit time across hundreds of very different species. The rates must also differ from protein to protein because some proteins such as cytochrome C have much wider variation across species than other proteins such as histone. Systemic macromutations of the biochemical system from mutations in a system-level genetic code (or other mechanisms) would arguably cause wide variations between all parts when a large jump occurs in the system and smaller variations between parts when a small jump occurs. This would reproduce the typological pattern often attributed to the hypothetical molecular clock.

The observed genetic code contains features that suggest a system-level genetic code. The genetic code for the blood clotting cascade contains similar sequences in different genes and within the same gene. These frequently appear to code for sequences in the proteins that bind to Vitamin K, another constituent of the blood clotting cascade[5,6]. This may be an example of a reusable standard interface component such as a connector. There are also adjacent pseudogenes that appear to be non-functional copies or near copies of the coding genes. Pseudogenes are a common part of the genetic structure. This may be similar to a master copy used to fabricate production dies but never used directly to manufacture the parts. This analogy would reverse the presumed order of gene duplication with the pseudogene acting as a precursor to the coding gene. The gene duplication transforms the non-coding pseudogene into the coding gene, several closely related coding genes, or homologous sequences within several coding genes. These correlated structures in proteins are usually attributed to random gene duplication and random shuffling of regions delimited by the introns. Regardless of their specific interpretation, correlations between seemingly independent parts would be expected if a system-level genetic code exists.

A system-level genetic code governing the morphology of living systems could resolve a major problem with homology. Homologous organs such as eyes frequently follow substantially different formation pathways during the development of the embryo[7]. The same organ or homologous structure may develop from completely different precursor structures in different species. This is extremely difficult to explain if the different parts such as bones are coded independently. This unstated assumption led early advocates of evolution such as Charles Darwin and Ernst Haeckel to expect similar development of embryos and homologous organs in different species. Similarly Haeckel's erroneous claim that "ontogeny recapitulates phylogeny" is a logical deduction if one assumes that each physical part of the adult organism is coded independently and therefore must develop essentially independently in the embryo. However, consider the toy example of the jigsaw puzzle coded as a series of rules for cutting a parent piece into matching smaller pieces. The order in which the puzzle is cut can be varied by swapping different cutting stages. If the order in which the cutting operations are executed varies, identical jigsaw puzzles can be produced by substantially different pathways.

System-level genetic codes are suggested by a variety of observations and theoretical considerations. The principal obstacle to a system-level genetic code theory is the apparent lack of any biochemical mechanism that implements a system-level genetic code. This objection is addressed below where it is demonstrated that biochemical mechanisms similar or identical to those observed in transposable elements, regulation of gene expression, and the adaptive immune system of jawed vertebrates can implement system-level genetic codes based on context-free grammars. It remains to be demonstrated whether the transposable elements and other features in the non-coding DNA implement a system-level genetic code. A system-level genetic code is feasible and a number of experimental observations including dramatic reorganizations of chromosomes mediated by transposable elements are consistent with the system-level genetic code hypothesis.

2. AN EXAMPLE DETERMINISTIC SYSTEM-LEVEL GENETIC CODE

It is possible to code for complex systems in such a way that all single mutations generate a complex system of co-adapted parts. For example, let the upper-case letters *A-Z* represent standard interfaces. The upper-case letter *A* represents a complementary pair of components *A-* and *A+* forming a standard interface. Examples of *A-* and *A+* are a lock and key, a plug and jack, a telephone transmitter and receiver, and two complementary interacting protein domains. Let the lower-case letters *a-z* represent components that play no role in the interface between parts.

| Symbol in Code | Meaning | Terminal Symbol |
| --- | --- | --- |
| (whitespace) | Separates Discrete Parts | Terminal |
| *A-, B-, C-, ...* | Negative Half of Interface | Terminal |
| *A+, B+, C+, ...* | Positive Half of Interface | Terminal |
| *a,b,c,...* | Attribute of a Part | Terminal |
| *A,B,C* .... | Standard Interface | Non-Terminal |
| S | System or Start Symbol | Non-Terminal |

This system-level code is a special case of a context-free grammar. A context-free grammar (CFG) is a concept used in compilers for computer languages and in linguistics. Compilers are programs that convert a program written in a high-level computer language such as C, C++, or FORTRAN to binary machine language instructions that a computer understands[8]. A context-free grammar consists of:

(1) A finite terminal vocabulary $V_t$

(2) A finite set of different, intermediate symbols, called the non-terminal vocabulary $V_n$

(3) A start symbol $S \in V_n$ that starts all derivations. A start symbol is sometimes called a goal symbol. The start symbol is a member of the set of non-terminal symbols.

(4) *P*, a finite set of productions, sometimes called production or rewriting rules, of the form $A \Rightarrow X_1 ... X_m$ where $A \in V_n$, $X_i \in V_n \cup V_t$, $1 \leq i \leq m$, $m \geq 0$

The context-free grammar starts with the start symbol. The production rules are applied until a series of only terminal symbols are reached. The production rules and the set of non-terminal symbols, the non-terminal vocabulary, form the system-level code. The final sequence of terminal symbols is the part-level code. Not all context-free grammars will constitute system-level codes. Conversely, some system-level codes may not be context-free grammars. Context-free grammars are relatively simple mathematical structures. Something more sophisticated may be needed to code living systems. In a hypothetical system-level genetic code the terminal symbols probably correspond to coding gene segments that encode functional domains within proteins, the exons found in most eukaryotes. The production rules may correspond to pseudogenes, transposable elements, or genetic modules that closely resemble transposable elements as explained below.

In the context-free grammar for this simple example system-level genetic code, the production rules for the context-free grammar specify a unique message rather than an allowed syntax of a family of messages. The message is encoded as a series of production rules for rewriting the system level code. These rules are iterated until a message formed of terminal symbols is reached. The system-level code is deterministic. The derivation of the part-level code is completely determined and always produces the same final part-level code, the final sequence of terminal symbols.

This simple example code has a single rewriting rule that translates a standard interface such as *A* into *A- A+* where *A-* and *A+* are separated by a white-space symbol. The white-space symbol may correspond to the start and stop codons in the known genetic code.

A cascade of interacting parts can be represented as a series of the letters. For example:

*aAabBaBcCa* (System-Level Code for a Chain of Five Coadapted Parts)

This is equivalent to the discrete parts:

*aA- A+abB- B+aB- B+cC- C+a* (Part-Level Code for a Chain of Five Coadapted Parts)

The system-level code acts as the master copy. The code translates the interface components *A-Z* into the two complementary parts: *A-* and *A+*, *B-* and *B+*, and so forth. Spaces delimit discrete parts in the part-level code and might correspond to start and stop codons in the DNA genetic code.

In the language of context-free grammars, the system-level code consists of four production rules. A colon is used to represent the rewriting operation:

(system) : *aAabBaBcCa*

*A : A- A+*

*B : B- B+*

*C: C- C+*

This system-level genetic code represents the system of discrete parts:

*aA- A+abB- B+aB- B+cC- C+a* (Part-Level Code for a Chain of Five Coadapted Parts)

Any single mutation in the system-level code such as addition, deletion, or change of any single letter to another letter in the alphabet yields another system of coadapted parts. For example, if the first upper-case *B* changes to a lower-case *b*, the code becomes:

*aAabbaBcCa* (System-Level Code for a Chain of Four Coadapted Parts)

This is equivalent to the discrete parts:

*aA- A+abbaB- B+cC- C+a* (Part-Level Code for a Chain of Four Coadapted Parts)

Any single mutation of the interfaces such as *A-* and *A+* in the part-level code will break the chain. There is also no way for a single mutation in the part-level code to bisect a piece into two complementary pieces preserving a chain of interacting parts. For example, *A+abbaB-* cannot be converted to *A+aB- B+baB-* by a single mutation in the part-level code. The mutations in the part-level genetic code discussed here correspond to the hypothetical random gene duplications, gene segment duplications, and gene segment shuffling currently invoked to explain the frequent sequence similarities in genes for distinct proteins.

The simple example allows some hard numbers to illustrate the difference between a system-level and part-level code. Consider the simplest system of coadapted parts, a system of two complementary parts:

*aAb* (System-Level Code for Two Coadapted Parts)

*aA- A+b* (Part-Level Code for Two Coadapted Parts)

Ignoring the two ends *a* and *b* for simplicity, there are twenty-five (25) single mutations of *A* yielding valid systems, for example *A -> B*. There are twenty-six (26) single mutations of *A* that yield invalid systems, for example *A -> b*. However, in the part-level code, there is no single mutation of the interface pieces *A-* or *A+* that yields a valid system. There are only twenty-five (25) double mutations that yield valid systems. There are 26*26 plus 25*24, a total of 1276, double mutations that break the system and yield invalid systems where the two parts will not work together. The probability of random variation in the part-level code improving the system is very small whereas the probability in the system-level code is actually quite high.

If the probability of a single mutation is one (1) per one-thousand (1000) generations, then the probability of a double mutation is one (1) per one-million (1,000,000) generations. Just as an example, suppose only one change represents a superior system, for example *aBb* also represented as *aB- B+b* is better than *aAb*. Organisms that contain the system coded by *aBb* are more able to survive and reproduce than those with *aAb*. For example, if *aAb* and *aBb* are parts of a primitive blood clotting cascade, then *aBb* might be less prone to accidental activation by proteins in the plaque forming on the inner surface of blood vessels, a speculated cause of heart failure. In the system-level code it would take about 26,000 generations to improve the species. In the part-level code it would take 1,000,000 generations on average to get one pair of simultaneous mutations. Of these, only 25 out of about 1301 double mutations are even valid systems. And only one (1) is the better case. It would take about 650,000,000 generations to make the one improvement, to find the better *aBb* case. The system-level code evolves about 20,000 times faster than the part-level code.

The single mutation rate of one per one-thousand generations must be significantly higher than the actual observed mutation rate for proteins in the blood clotting cascade. If this were true, one in a thousand children would be born with usually deadly blood disorders. The incidence of hemophilia and other congenital blood diseases would limit the rate to something more like 1 in 3,500 generations even if all cases of hemophilia and other diseases were attributed to a current mutation instead of one inherited from past generations.

The *aAabBaBcCa* system-level genetic code corresponds to a rigorous top-down design in which a designer partitions a system into black boxes. Standard interfaces are selected and used to specify the relationship between the different black boxes. For example, a stereo system designer selects the standard RCA stereo jack and plug as the interface between the different modules in the stereo system such as receiver, amplifier, and speakers. The standard interfaces such as the RCA jack and plug correspond to the upper-case letters *A-Z* in the code. Each black box in the design has attributes that are independent of the interfaces specified. These attributes can be changed without modifying the interfaces. In a stereo this might correspond to the choice of a vacuum tube based FM tuner or a transistor-based tuner. These attributes correspond to the lower-case letters *a-z* in the code. For example, *a* might code for a vacuum tube based FM tuner. The lower-case letter *b* might code for a transistor-based FM tuner.

The human designer achieves great efficiency by choosing the necessary interfaces for the parts from a library of pre-existing interfaces, coded by *A-Z*, and a library of pre-existing attributes, coded by *a-z*.

The usefulness of the code should be enhanced by adding a mechanism for representing functional modules such as chromosomes or segments of chromosomes. Symbols and production rules representing large genetic modules are significant because chromosomal rearrangements are seen in cancer cells in the body, in immortal "normal" and tumor cell lines, in maize through the action of transposable elements, and in a number of other contexts. This can be achieved by adding non-terminal symbols $\alpha$–$\omega$ to represent the hypothetical genetic modules.

| Symbol in Code | Meaning | Terminal Symbol |
|---|---|---|
| (whitespace) | Separates Discrete Parts | Terminal |
| *A-, B-, C-, ...* | Negative Half of Interface | Terminal |
| *A+, B+, C+, ...* | Positive Half of Interface | Terminal |
| *a,b,c,...* | Attribute of a Part | Terminal |
| *A,B,C ....* | Standard Interface | Non-Terminal |
| *α,β,λ, ...* | Functional Modules | Non-Terminal |
| *S* | System or Start Symbol | Non-Terminal |

In this code, a complex system can be coded as a hierarchy of production rules. For example:

*S : αβ*

*α : aAbbBcdCa*

*β : aCddCa*

*A : A- A+*

*B : B- B+*

*C: C- C+*

This codes the final system of discrete parts:

*aA- A+bbB- B+cdC- C+aaC- C+ddC- C+a*

If copying during cell division or some other genetic process proceeds in a modular fashion treating symbols in the system-level genetic code as units, copying errors can now duplicate, delete, swap, or change the symbols α and β for functional modules, producing large-scale restructuring of the system. Thus it is demonstrated that mathematical systems exist that correspond to system-level genetic codes as defined in this paper. These systems can produce large-scale coordinated changes analagous to the hypothetical systemic macromutations required for saltational theories of evolution and for chromosomal mutation theories of cancer.

The obvious problem with the system-level genetic code theory is the apparent lack of a plausible biochemical mechanism to implement the system-level genetic code. Transposable elements are genetic systems that move or copy themselves to other locations within the genome. They have proven ubiquitous in genetic systems. Since their discovery transposable elements have been proposed as controlling elements in the genetic system that mediate large scale genetic changes[9,10,11]. The dominant view is that transposable elements are genetic parasites that coexist with the coding DNA but serve no useful purpose. The example system-level genetic code in this section can be implemented using molecular machinery similar or identical to that observed in transposable elements, regulation of gene expression, and the adaptive immune system of jawed vertebrates.

One can implement a production rule in a context-free grammar as a transposable element-like genetic module. Consider, for example, the first production rule in the simple example system-level genetic code:

*S* : αβ

This can be implemented in the DNA sequence as a transposable element-like genetic module:

(Transposable Element Regulatory Binding Site) …. (Transposition Start Code)(DNA sequence for a transposase that targets a recombination signal *S*)(Inactive Recombination Signal α)(Inactive Recombination Signal β)(Transposition Stop Code)

The Transposable Element Regulatory Binding Site functions as an on/off switch that activates the transposable element. Under normal conditions, the transposable element is dormant. This corresponds to the observed behavior of transposable elements. The transposable elements are normally dormant except in response to stresses such as starvation.

When the transposable element is activated by the cellular regulatory system in response to an environmental stress such as starvation or a carcinogen the transposase is synthesized. The transposase is a special protein that can move or copy segments of DNA. This transposase will cut at the transposition start code, making a copy of the codes delimited by the transposition start and stop codes including the code for the transposase. This DNA is bound with the transposase into a transposase-DNA complex. Like various regulatory proteins, the transposase binds only to specific binding sites in the DNA, recombination signals, that identify targets for a move or copying operation. The transposase-DNA complex floats around in the nucleus until it finds an active recombination signal *S* or it breaks down. The recombination signals correspond to the non-terminal symbols in the system-level genetic code.

When the transposase-DNA complex finds the targeted recombination signal, such as the non-terminal symbol *S* in the example, the transposase replaces the active recombination signal *S* with the codes to the "right" of the DNA sequence for the transposase – for example, the recombination signals α and β in the example. The codes to the "right" of the transposase can also include inactive non-coding forms of standard gene segments such as *a-z*, *A-* ,and *A+* in the example.

To avoid overwriting the system-level genetic code itself the recombination signals and prototype gene segments in the production rules are inactive or non-coding. The code for the recombination signals is modified to insure that an active transposable element will not bind to the recombination signals in the DNA for the system-level genetic code.

However, when the active transposable element replaces an active recombination signal such as *S*, the recombination signals or gene segments that replace *S* are activated.

When activated recombination signals, the non-terminal symbols in the system-level genetic code, are present in the genome, the cellular regulatory network generates a regulatory protein or proteins that binds to the regulatory binding sites of the transposable elements comprising the system-level genetic code. This protein or proteins activates production of the transposable elements. Production of the transposable elements continues until all active recombination signals have been deleted, until all non-terminal symbols have been rewritten as terminal symbols.

To summarize, re-derivation of the part-level genetic code is triggered by a stress that causes the cellular regulatory network to insert an activated start symbol *S* somewhere in the genome. This stress could be starvation or a chemical toxin such as a carcinogen. Once the process is initiated, the cellular regulatory network generates a protein or proteins that cause production of the transposable elements. Activated non-terminal symbols cause the cellular regulatory network to generate the transposable element activation protein or proteins. Each transposable element will replace an active non-terminal symbol with a sequence of symbols. The active non-terminal symbols are recombination signals that the transposase in the transposable element binds to. This process continues until there are no activated non-terminal symbols. When the activated non-terminal symbols are consumed, the cellular regulatory network ceases production of the activating protein or proteins. The activated terminal symbols correspond to the coding gene segments, the exons.

Because re-derivation of the part-level genetic code may involve global restructuring of the genome such as duplications of chromosomes, the re-derivation probably occurs during cell division when the chromosomes are unraveled.

In the model presented above, the symbols in the production rules forming the system-level genetic code need to be inactivated to prevent the re-derivation process from overwriting the system-level genetic code. Non-coding sequences such as the pseudogenes are therefore necessary and expected in this model. The non-coding sequences may closely resemble the coding sequences. This may be a general feature of system-level genetic codes.

3.  AN EXAMPLE STOCHASTIC SYSTEM-LEVEL GENETIC CODE

The previous example is a deterministic system-level genetic code. The code uniquely specifies a single system of parts. Variations will be produced only by accidental mutations such as copying errors. It is speculated that copying during cell-division is modular, acting on the symbols as units, so that mutations such as duplications, swapping of symbols, and so forth are probable. The example is deterministic because a single production rule exists mapping a non-terminal symbol to a sequence of symbols. Each right hand side in the context-free grammar is unique. No mechanism for generating a diversity of systems exists. System diversity may be needed to generate the extreme complexity of higher organisms such as the human brain. In addition, a mechanism to generate system diversity would be useful for any adaptive mutation mechanism.

The system-level genetic code can be extended to support system diversity. These system-level genetic codes will be called stochastic system-level genetic codes in this paper.

For example:

*S* : *αβ*

*α* : *aAbbBcdCa*

*α* : *aAbbCa*

*β* : *aCddCa*

*A* : *A- A+*

*B* : *B- B+*

*C*: *C- C+*

This system level code represents two systems:

*aA- A+bbB- B+cdC- C+aaC- C+ddC- C+a*

and

*aA- A+bbC- C+aaC- C+ddC- C+a*

Using a transposable element-like mechanism as before, there are now two transposable elements with a transposase binding to the recombination signal represented by α instead of one. There is an equal likelihood that one or the other will be used in generating the final part-level genetic code.

Consider a stochastic system-level genetic code in which ten non-terminal symbols (α,β,χ,δ,ε,φ,γ,η,ι,φ,κ) each have two distinct equally likely production rules. This system-level genetic code can represent $2^{10}$, about one-thousand, different systems of parts. A small increase in the number of production rules can increase the number of systems represented or accessible to the code exponentially.

*S* : *αβχδεφγηιφκ*

*α* : *aAbbBa*

*α* : *aAbB*

*β : bbAbb*

*β : bcAbCb*

...

*κ : aaBccBc*

*κ : aaAcBc*

*A : A- A+*

*B : B- B+*

*C : C- C+*

One-thousand and twenty-four (1,024) different systems are represented by twenty-four (24) production rules. Each production rule corresponds to a transposable element-like genetic module. Each non-terminal symbol corresponds to a recombination signal, a binding site in the DNA.

A stochastic system-level genetic code can compactly represent a huge diversity of distinct but related cell types such as the many types of neurons and glial cells in the brain. It provides a plausible mechanism for extremely complex cell differentiation in multi-cellular organisms.

A stochastic system-level genetic code provides a mechanism for sophisticated somatic (somatic refers to cells in the soma or body of an organism) adaptation to environmental stresses since a dividing cell can mutate into a very different type of cell through stochastic re-derivation of the part-level genetic code. The stochastic re-derivation process can generate entirely new systems of proteins or regulatory networks in a single cell division. Thus, the cell can adapt to stresses that could not be handled by changing or creating a single protein as in the adaptive immune system of jawed vertebrates. More sophisticated mechanisms could pass these adaptations to other cells and to future generations as in the transfer of antibiotic immunity genes by plasmids in bacteria.

The usefulness of a stochastic system-level genetic code would be greatly enhanced if a feedback mechanism to copy useful adaptations in somatic cells into the system-level genetic code of the germline (germline refers to reproductive cells such as ova and sperm) cells exists. In this case, a useful adaptation would be passed on to future generation. Endogenous retroviruses provide a plausible mechanism to implement the needed feedback if they can capture messenger RNA (mRNA) segments corresponding to production rules in the system level genetic code or genes in the conventional genetic code and integrate these segments into their RNA-based genomes[12]. The endogenous retroviruses could also pass acquired characters laterally to other animals.

Under certain stresses maize undergoes dramatic chromosomal rearrangements mediated by transposable elements. These vary from cell to cell in the developing plant so that different parts of the plant develop different features such as coloration patterns in leaves. Seeds that develop from different parts of the plant pass these traits on to offspring. Maize does not require a complex feedback mechanism such as endogenous super-retroviruses to pass on somatic changes to daughter plants. The behavior of transposable elements in maize strongly resembles a hypothetical stochastic system-level genetic code.

4. A STOCHASTIC SYSTEM-LEVEL GENETIC CODE WITH MEMORY

A problem with the simple stochastic system-level genetic code is how does it evolve since the same code represents both adaptive and maladaptive systems. Any adaptive changes will apparently be lost the next time that the part-level genetic code is re-derived. A simple feedback mechanism permits a stochastic system-level genetic code using transposable elements to learn from successful adaptations. So far the discussion has assumed that the active transposase-DNA complex is discarded after replacing the left hand side of a production rule, a single non-terminal symbol encoded by a recombination signal, with the right hand side. If instead the transposase-DNA complex that *is used* during re-derivation is inserted into the DNA as a dormant extra copy of the transposable element, then the number of copies of the production rule used during re-derivation is increased. Transposase-DNA complexes that *are not used* are discarded and not inserted into the DNA as dormant transposable elements.

Cells that are maladaptive perish. Adaptive cells contain multiple copies of the production rule that led to successful adaptation. These cells are then more likely to use the adaptive production rule during the next re-derivation of the part-level genetic code. In a simple scheme, when re-derivation is triggered, all dormant transposable elements generate transposase-DNA complexes. A transposable element with many duplicate copies in the genome generates many transposase-DNA complexes. A transposable element with only one copy in the genome generates a single transposase-DNA complex.

Repeated re-derivations of the part-level genetic code using this simple mechanism with cumulative transposable elements will result in numerous repeated copies of adaptive transposable elements that implement adaptive production rules. This may explain the long term repeating sequences in the non-coding DNA that often appear to be repeated copies of the same transposable element.

Feedback from somatic to germline cells is not necessary for a stochastic system-level genetic code to learn from successful adaptations. The accumulation mechanism can be restricted to germline cells. Organisms that survive will pass repeated copies of adaptive transposable elements on to their offspring. Adaptive transposable elements will accumulate over many generations – each time the part-level genetic code is rederived from the stochastic system-level genetic code.

The stochastic system-level genetic code with memory will accumulate a rough count of the number of times a production rule was used in ancestors *that survived*. In turn, this historical record determines the probability that the production rule is used when the part-level genetic code is rederived.

For example, consider two alternative production rules:

$\alpha : aAbbBb$   (75 copies in genome)

$\alpha: aAb$      (25 copies in genome)

This means that ancestors that survived used the first production 75 % of the time and used the second production rule 25 % of the time.

It is a bad strategy to always use the first production rule. For example, the first production rule might be optimum during a normal climate but the second production rule might represent an optimal feature during a drought. Droughts occur randomly one out of four years in the organism's habitat. If all organisms of the species use the first production rule, the entire species will perish during the drought. A better strategy for survival is to use the first production rule 75 % of the time and the second production rule 25 % of the time as in this example.

More sophisticated memory and learning algorithms for stochastic system-level genetic codes are possible. For example, a mechanism to weight recent adaptive production rules more strongly than past production rules would probably enhance the adaptive performance of the code. This could be accomplished simply by having a transposable element that was used during re-derivation make multiple copies of itself instead of a single copy and an unused transposable element delete extra copies of itself, preserving some copies for extreme conditions where it was adaptive.

5. THE REGULATORY BINDING SITES AND THE SYSTEM-LEVEL GENETIC CODE RECOMBINATION SIGNALS MAY BE THE SAME OR CLOSELY RELATED

In the system-level genetic codes presented here, a genetic sub-system is identified by recombination signals that are binding sites for the transposases or other constituents of transposable element-like modules in the genetic system. In real living organisms, these genetic systems must be subject to coordinated regulation by regulatory proteins generated by regulatory genes that control gene expressions. Thus, genes for interacting proteins frequently share common regulatory binding sites upstream of the coding gene that allow a single regulatory protein to simultaneously control several distinct genes.

The simple models in this paper do not explain how the system-level genetic codes could generate the complex regulatory networks for gene expressions. These must be generated simultaneously with the coding genes for the structural proteins. This problem lies beyond the scope of the current presentation. One important possibility is worth mentioning.

Considerable efficiency and robustness to errors can be achieved by using the same binding sites as targets for both the regulatory proteins and the hypothetical transposable elements. Alternatively, the binding sites for the regulatory proteins may be generated from the binding sites for the transposable elements by a simple conversion. During the rewriting operation the non-terminal symbol, the recombination signal in the DNA, on the left hand side of the production rule may not be discarded. Instead the recombination signal is copied or converted to a regulatory binding site for each symbol on the right hand side of the production rule. Thus, all components of a sub-system will be regulated by the same regulatory protein or proteins. A tight coordination between systems of coadapted proteins as identified by the system-level genetic code (which insures that the different proteins fit together as in the blood coagulation cascade) and the systems as identified by the regulatory networks that control gene expression is ensured. In general, the regulatory networks that control gene expression will need to turn on or off the proteins in the sub-systems of interacting proteins defined by the system-level genetic code at the same time. Otherwise some proteins in the sub-system will not be expressed and the sub-system will not function.

This dual use of binding sites may also explain why transposable elements are often observed to contain what appear to be codes used in the regulation of gene expression. These codes perform a dual role, governing gene expression and directing the assembly of networks of coadapted structural proteins.

## 6. CANCER AND CELL LINES

Cancer is the second leading cause of death in industrial nations. There is about one chance in five that the reader will contract and die from cancer. In most cases, if the tumor cannot be surgically removed before it spreads, the tumor will eventually kill the patient.

Cancer is a good candidate for a systemic macromutation of a single cell. In general, mutations able to produce detectable changes in organisms are negative mutations resulting in organisms that soon perish or would perish in the wild. Most random mutations of single cells in organisms should result in cells that simply die due to a malfunction of the complex cellular machinery. Cancer cells on the other hand become unusually robust. Most strains that have been cultured in the laboratory become immortal and can survive without the supporting machinery of the organism[13]. Cancer cells frequently exhibit large-scale chromosomal abnormalities that suggest a large-scale restructuring of the genome[14,15]. This is difficult to reconcile with point-mutation, viral oncogene, oncogene, and anti-oncogene hypotheses. However, a mutation in the hypothetical system-level genetic code might produce these large changes in the chromosomes.

It is striking that chromosomal abnormalities exist in cancer cells at all. Random rearrangement of man-made machines – imagine, for example, interchanging a car engine and back seat or, worse, interchanging half of a car engine and half of a back seat – invariably renders the machines non-functional and it is difficult to see how random changes to chromosomes would not produce catastrophic results. In contrast, the point-mutation, oncogene, viral oncogene, and anti-oncogene hypotheses would be similar to the car ignition switch breaking so that the car cannot be turned off – a simple, plausible localized mutation of a single part or a few parts that could cause cancer without destroying the cell. Yet, contrary to naive intuition, cancer cells frequently contain demonstrable chromosomal abnormalities. This strongly suggests that the chromosomal abnormalities frequently found in cancer cells are not random changes.

The transition from single-cell to multi-cellular organisms required the creation of a control system that forced the cells to cooperate. The system-level genetic code theory attributes this transition to a mutation in the system-level genetic code that created the control system in a single step or a small number of steps. Cancer would be the obvious result if this control system failed, disappeared, or substantially changed due to a systemic macromutation in a single cell. A large-scale change in chromosomal structure, whether caused by a mutation in the hypothetical system-level genetic code or not, could easily overwrite or disable the entire multi-cellular control system. The cell would revert to an autonomous single-celled organism and devour its host. In the system-level genetic code theory, the instructions for the multicellular control system are contained in the system-level genetic code and it is in the system-level code that the change or changes that cause cancer occur. Cancer is a leap back across the evolutionary divide between multi-cellular organisms and the primordial single-celled organisms.

If cancer were attributed to the failure or deactivation of a regulatory gene or genes governing the multi-cellular control system then the large-scale chromosomal abnormalities frequently observed in cancer cells would not occur. The cancer cell would differ from healthy cells by only one or a few regulatory genes. Further, mapping of these small differences using gene sequencing technology would have rapidly identified the genetic and biochemical basis of cancer through comparison of cancer cells and healthy cells. The regulatory genes would necessarily code for proteins that would be damaged or absent in the cancer cells. Cancer could then be treated simply by synthesizing the protein manufactured by the undamaged genes in the healthy cells and flooding the tumor with the correct regulatory proteins. This would restore the multi-cellular control system to healthy function inhibiting or even eliminating the tumor. This has obviously not been achieved – probably because the cancer cells differ dramatically from the healthy cells at a genetic level and the change or changes that cause cancer are difficult to identify.

If the hypothetical system-level genetic code is modified, then the entire part-level genetic code may be re-derived from the system-level description. Chromosomes could be rearranged, not in a random manner that would almost certainly produce a non-functional cell but in a highly organized manner producing a highly functional killer cell that devours its host. For example, the sections in the chromosomes rearranged in the chromosomal abnormalities are not selected at random but correspond to the genetic code for an entire sub-system such as the multi-cellular control system. The simplest explanation is that the rearrangements shut down the multi-cellular control system or a critical part of the multi-cellular control system.

If re-derivation of the part-level genetic code, the coding genes, from the system-level genetic code occurs, the re-derivation probably only occurs during cell division. Since the re-derivation can cause global systemic changes in the DNA, such as for example restructuring of chromosomes, the entire DNA must be unraveled into separate strands as occurs during cell division. The partial unraveling of small regions of the DNA double helix when the messenger RNA for isolated genes is produced will, in general, be inadequate for global changes. There is no requirement that the re-derivation of the part-level genetic code occurs during every cell division. Indeed it may occur only under specialized conditions.

In the system-level genetic code theory of cancer suggested here, the multi-cellular control system is either absent or substantially modified in cancer cells. An entire system of proteins is either absent from or greatly changed in cancer cells. This makes treating or curing cancer more difficult than a cancer due to the few defective or missing genes in conventional genetic theory. Simply replacing a defective or absent protein will not work because the proteins that react with this protein are also missing from the cancer cells. On the other hand the cancer cells will contain substantial differences from the healthy cells since an entire biochemical system is either absent or grossly modified. A "magic bullet" protein that attacks only the cancer cells should be easier to develop once the biochemical system has been identified, possibly through decoding of the system-level code for the multi-cellular control system. What is needed in this case is a destructive protein that is easily deactivated by a multi-cellular control system protein found only in the healthy cells or a proenzyme that is activated by a mutant multi-cellular control system protein found only in the cancer cells. The proenzyme is converted to a destructive enzyme that destroys the cancer cell. The proenzyme method will not work if the multi-cellular control system is absent in the cancer cells. The proenzyme might be created by crossing a digestive enzyme with a constituent of the mutant multi-cellular control system from the cancer cells.

If the cancer cell lacks one or more proteins found in healthy cells – for example, the multi-cellular control system or a significant part of the control system is missing – the "magic bullet" could be fashioned from a cascade of two or more tightly coupled proteins. With two proteins, the simplest "magic bullet", the first protein is a proenzyme for a destructive enzyme that will kill the cell such as a digestive enzyme or biological toxin. The second enzyme would be an activation enzyme for the proenzyme. The activation enzyme is engineered so that a regulatory protein found only in the healthy cells destroys the activation enzyme. Treatment would consist of first introducing the activation enzyme into the patient. The activation enzyme would be destroyed in the healthy cells but build up to a relatively high concentration in the cancer cells. Ideally the activation enzyme should affect only the proenzyme to avoid side-effects. Next, the "magic bullet" proenzyme would be introduced into the patient. The proenzyme would remain inactive in the healthy cells but would be converted to the enzyme in the cancer cells that now contain relatively high concentrations of the activation enzyme. Because the concentration of the multi-cellular control system proteins is probably quite low in both the healthy and cancer cells, the cascade may need more than two proteins to amplify the signal. The building blocks of the "magic bullet" cascade can be fashioned by cannibalizing parts of the multi-cellular control system cascade from healthy cells.

Something more sophisticated will be needed if the changes in the cancer cells change the relative frequencies of proteins or the temporal order of production of proteins in the multi-cellular control system but do not delete or modify any of the proteins or add new proteins that indirectly modify the system. For example, the changes could duplicate genes causing excess production of regulatory proteins. Alternatively, the position of the genes along the chromosome or on different chromosomes or relative to other genes or non-coding markers could control the temporal order in which proteins are synthesized. Nonetheless, these speculations illustrate how decoding the system-level genetic code if it exists could lead directly to a treatment for cancer.

Any theory attributing cancer to a failure of the multi-cellular control system can explain the difficulty in finding the biochemical difference between cancer cells and healthy cells, the presumably missing or damaged proteins, needed for effective treatment. Control systems such as the early stages of the blood clotting cascade are information processing systems and do not require large, easily detectable concentrations of proteins to function. They act as triggers turning on or off blood clotting, cell division, and other functions. Thus it might be virtually impossible to identify the missing or damaged proteins from chemical analysis and comparison of cancer cell cultures and healthy cell cultures. If the damaged genes or the entire system of genes could be identified, the genes from healthy cells and cancer cells could be inserted into cooperative bacteria through recombinant DNA techniques and the missing or damaged proteins mass-produced for research and treatment purposes.

A system-level genetic code using a transposable element-like mechanism could resolve most of the paradoxes of cancer. Many carcinogens are known that cause cancer but do not directly damage the DNA. In a transposable element system-level genetic code, the re-derivation of the part-level genetic code is triggered by regulatory proteins that activate the transposable element-like constituents of the genetic system. The regulatory proteins could easily be affected by carcinogens that attack proteins but not the DNA. Direct damage to the DNA could also generate an active non-terminal symbol triggering partial or complete re-derivation of the part-level genetic code. Thus, both non-mutagenic carcinogens and DNA damaging mutagens can plausibly cause cancer if a system-level genetic code exists.

Cancer tends to occur in dividing cells and rarely in non-dividing cells. Re-derivation of the part-level genetic code is a global phenomenon that would require unraveling of all chromosomes. It probably occurs mostly during cell division.

Cancer cells almost always have large-scale chromosomal changes such as duplications, translocations, and inversions of chromosomes as discussed above. The chromosomal abnormalities are the most consistent marker of cancer cells. A mutation in a high-level production rule of a system-level genetic code could easily produce large scale restructuring of chromosomes that would not be fatal to the cell.

Cancer tends to occur in middle and old age. The body is beginning to break down due to the aging process. The cells are subjected to various stresses that do not occur during youth. Thus, triggering of the re-derivation process may be more likely in an aging body.

The extent of the chromosomal abnormalities in cancer cells is positively correlated with the aggressiveness and lack of differentiation of cancer cells. The more aggressive, the more prone to spreading through the body, and the less differentiated the cancer cell, in general the more extensive the chromosomal abnormalities. If the chromosomal abnormalities are interpreted as a systemic macromutation, a leap back across the evolutionary divide toward a unicellular organism, this relationship is easy to understand. The more extensive the reorganization of the chromosomes, the further the cell has leapt back toward a unicellular organism.

Non-cancerous chromosomal abnormalities have been observed in the elderly and in survivors of the atomic bombs in Hiroshima and Nagasaki. Usually, the macromutations caused by the mutations in the system-level genetic code generate systems of parts, new types of cells, that are neither cancerous nor defective. These cells are examples of "successful" somatic macromutations that do not threaten the host organism. These mutations tend to be produced by stresses or genetic damage to the system-level genetic code.

The imperfect correlation of cancer with various supposed oncogenes can be explained if the system-level genetic code also modifies individual coding genes during the re-derivation process. This can be caused by shuffling of exons since individual coding genes are constructed from gene segments in a system-level genetic code or by small variations similar to the somatic hypermutation phenomenon observed in the adaptive immune system of jawed vertebrates. In this latter hypothesis, the transposases make random or directed modifications to gene segments that they copy. These changes are more extensive than converting a non-coding gene segment into a coding gene segment, an exon. In a stochastic system-level genetic code, these small changes to individual coding genes are randomly produced and hence are seen in only some cancer cells. The creation of oncogenes is probably peripheral to the major systemic changes that invariably cause cancer since the oncogenes are often not found in cancer cells.

Oncogenes may be produced by "deliberate" small variations or errors in joining gene segments into coding genes. Not all non-terminal symbols in the system-level genetic code may correspond directly to exons. Some symbols could be building blocks used to assemble exons. Thus, even oncogenes that occur within an exon or within a DNA sequence that appears to correspond to a single folding domain in a protein may derive from joining errors not unlike the variations that occur during assembly of antibody genes in the T and B cells.

Supposedly oncogenic (cancer-causing) retroviruses that are not immediately transforming but cause cancer in laboratory animals after a long latent period can be explained within this theory. The slow oncogenic retroviruses may either carry an oncogenic production rule within the system-level genetic code that is inserted in the cells of the animal or they cause mutations of the system-level genetic code during insertion into "non-coding" DNA that carries the system-level genetic code. In both cases cancer will only occur when the part-level genetic code is re-derived which requires a subsequent stress, months or years later. There is a long, unpredictable latent period until re-derivation. This also explains why cancers attributed to infection with slow-acting retroviruses always have chromosomal abnormalities and are monoclonal (only one cell turns into a cancer cell and all tumor cells are derived from this one cell) unlike immediately transforming retroviruses such as the Peyton Rous sarcoma virus. Most cells infected by the retrovirus never undergo the re-derivation process. Further, in a stochastic system-level genetic code with memory, the oncogenic production rule would be infrequently used during re-derivation compared to the frequent copies of the adaptive non-oncogenic production rules. Thus, only a single infected cell undergoes transformation to a cancer cell. The animal (or person) usually dies from the tumor before a second cell undergoes the rare transformation to a cancer cell.

If a feedback mechanism to the germline cells exists, the association of cancer with retroviruses and the origin of the oncogenic retroviruses can be easily explained. The cancer cells do not realize that they are defective. The cancer cells assume they have succeeded in generating a viable macromutation that should be passed on to future generations. The cancer cells begin to produce retroviruses in an attempt to transmit their improvement to the host organism and possibly to other organisms of the same or closely related species. In some cases the re-derivation of the part-level genetic code manufactures a singular oncogene that can transform normal cells into cancer cells such as the oncogene in the Peyton Rous sarcoma virus. The oncogene is then integrated into an endogenous retrovirus to transmit the gene through the blood to the germline cells and laterally to other members of the same species.

The remarkable properties of both normal and tumor cell lines are easily accounted for by a system-level genetic code, especially a stochastic system-level genetic code. Both normal and cancer cells that are removed from the body and cultured in Petri dishes undergo dramatic changes. These cell lines frequently become immortal. Extensive chromosomal rearrangements are observed. Sex chromosomes are frequently discarded. These chromosomal changes frequently continue as long as the cell line lives in vitro. The cells usually loose most of their visible differentiation[16]. Most significantly, the immortal cancer cell lines often behave differently from the cancer cells in the body.

Once a cell is transplanted to a Petri dish, it is in a stressful environment where it cannot survive in its current form. To survive, it activates the system-level genetic code in an attempt to adapt. A stochastic system-level genetic code, in particular, can produce a vast number of different new cell types during cell division, most of which will perish in the hostile environment of the Petri dish. Even a deterministic system-level genetic code is subject to mutations due to accidental or preprogrammed copying errors. These mutations in the system-level genetic code cause the numerous and on-going chromosomal abnormalities seen in the cell lines. In the stressful environment of the Petri dish, the cell continues to mutate radically, something it has no need to do in the body. The cancer cells in the body believe that they have successfully adapted to the stresses in the aging body and remain stable. The regulatory networks in the body tend to suppress activation of the system-level genetic code. The cell lines in vitro, however, rapidly evolve and continue evolving into new types of single-celled organisms able to survive in the hostile environment of the Petri dish. Sex chromosomes are frequently discarded because the unicellular organisms no longer need the sexual traits required by multicellular organisms.

7. AN EXPLANATION FOR THE ABILITY TO REVERT TO UNICELLULARITY

Since cancer is an ultimately fatal disease in most cases, the question remains why evolution and natural selection have not blocked off the genetic route back to unicellularity indicated by cancer cells in the body and cell lines in vitro. Why might the ability to revert to unicellularity be useful?

There is strong evidence that the Earth and all life on Earth have suffered several global catastrophes. The most well-known is the apparent asteroid collision thought to have killed the dinosaurs and many other species at the end of the Cretaceous era. Several other mass extinctions, currently attributed to asteroid collisions, are known. In the extreme, an asteroid collision may have filled the Earth's atmosphere with dust obstructing light from the Sun, terminating all photosynthesis and killing all plant life. In this case, the production of oxygen would have ceased and the Earth would have become uninhabitable by nearly all multicellular organisms. Under such conditions, possibly only single-celled organisms could survive either by tapping energy sources not dependent on sunlight or by hibernation until the dust settled out of the atmosphere.

If the mass extinction at the end of the Cretaceous involved such a doomsday scenario, one is left with the enigma of how any animals or plants survived the hypothetical catastrophe. However, if some of the cells in the dying plants and animals could revert to single-celled organisms that could survive in the hostile environment, the enigma is resolved. The single-celled organisms would have retained all or most of the genetic information for functional animals, dormant until the environment reverted to an environment hospitable to multicellular life. Once the habitable environment was restored, the single-celled organisms could have re-evolved back into animals in a geologically short period of time.

Of course, a reversion to undifferentiated single-celled existence is precisely what appears to be observed in animal cells cultured in Petri dishes. This may be exactly the behavior needed to survive the catastrophes suggested by the fossil record. Hence, also, the widespread appearance of cancer in old age when the body is dying or in response to environmental shocks such as toxic carcinogens. Cancer may thus be an evolutionary adaptation of life to ensure survival under extreme conditions.

8. CONCLUSIONS

System-level genetic codes may account for a variety of problems with standard evolutionary theory and probably can provide a mechanism for systemic macromutations without introducing an intelligent agent, vitalistic forces, or new laws of physics. System-level genetic codes may be the fundamental causal mechanism underlying cancer in the body and the development of immortal "normal" and cancer cell lines in vitro.

The ideas in this paper strongly suggest a thorough reanalysis of the non-coding DNA, including the pseudogenes and the introns, especially non-coding DNA associated with the coding genes for complex integrative biochemical systems such as the blood clotting cascade or the adaptive immune system of jawed vertebrates. The DNA associated with complex biochemical systems is most likely to contain the system-level codes if a system-level genetic code exists. Context-free grammars, attribute grammars, and related concepts from computer science and linguistics are good candidates for the underlying mathematical structure that should be sought in the non-coding DNA. Transposable elements appear to have the properties necessary to implement production rules for a context-free grammar and perhaps more sophisticated formal languages.

The example system-level genetic code using transposable element-like mechanisms is a proof of concept. It is clearly too simple to implement the complex feedback loops and regulatory networks governing gene expression in real organisms. A more sophisticated system-level genetic code is needed. This may be based on context-free grammars. It seems likely that context-free grammars are too simple. A related, more advanced mathematical structure will probably be needed.

New experiments may be needed to verify or falsify the system-level genetic code hypothesis. However, vast quantities of gene sequencing data and other molecular biology data have been collected in recent decades. All the necessary data to decipher the system-level genetic code may already be in hand. The current situation may be like the state of Egyptology prior to discovery of the Rosetta stone. Extensive samples of hieroglyphics existed that no one could read. Additional hieroglyphics were of no use. The key lay in finding the Rosetta stone that provided translations of enough hieroglyphics into known languages to decipher the rest of the language from context. The Rosetta stone for the system-level genetic code may be found by careful reanalysis of existing data or a critical new experiment. The practical benefits of deciphering the system-level genetic code should be extensive.

9. ACKNOWLEDGEMENTS

This paper benefitted from discussions and correspondence with Hal Cox. It also benefited considerably from reading a preprint of a paper by James Shapiro of the University of Chicago to appear in *Genetica* : "Transposable elements as the key to a 21$^{st}$ Century view of evolution". It also benefitted from discussions with Ray Cowan. Of course, all errors and mistakes are the fault of the author. The author's background is in data compression and physics. Apologies are extended in advance to molecular biologists, biochemists, and others whose previous work on similar or identical ideas has not been cited.

10. ABOUT THE AUTHOR

John F. McGowan is an engineer and researcher in the field of digital video at NASA Ames Research Center. He is Technical Lead for the Desktop Video Expert Center at NASA Ames. He has worked on digital video quality metrics, still image quality metrics, perceptual optimization of JPEG still image compression, and the conceptual design of a video system for the Mars Airplane. He has designed and implemented MPEG-1 and MPEG-2 digital audio and video software decoders. He is the author of John McGowan's AVI Overview, a popular Internet Frequently Asked Questions (FAQ) on the AVI digital audio/video file format. He worked on the Mark III and SLD experiments at the Stanford Linear Accelerator Center (SLAC). John F. McGowan has a B.S. in Physics from the California Institute of Technology and a Ph.D. in Physics from the University of Illinois at Urbana-Champaign. His personal web page is http://www.jmcgowan.com/. He can be reached at jmcgowan@mail.arc.nasa.gov or jmcgowan@veriomail.com.